\documentclass[aps,prl,twocolumn,groupedaddress,amsmath,superscriptaddress]{revtex4-2}

\usepackage{amsfonts}
\usepackage{amsmath}
\usepackage[dvips]{graphicx}
\usepackage{color}
\usepackage{amsmath}
\usepackage{graphicx}
\usepackage{amssymb}
\usepackage{hyperref}
\usepackage{color}
\usepackage{mathrsfs}
\usepackage{isomath}
\usepackage{amsmath}
\usepackage{amsthm}
\usepackage{epstopdf}
\usepackage{txfonts}
\usepackage{dsfont}
\usepackage{setspace}
\usepackage{multirow}
\usepackage{booktabs}
\usepackage{tabularx}
\usepackage{colortbl}
\usepackage{hhline}
\usepackage{xcolor}
\usepackage[normalem]{ulem}
\allowdisplaybreaks[4]

\makeatletter
\definecolor{nblue}{rgb}{0.3,0.3,1.0}%229
\definecolor{ngreen}{rgb}{0.2,0.7,0.2}%161
\definecolor{nred}{rgb}{0.9,0.1,0}%711&900
\definecolor{nblack}{rgb}{0,0,0}

\newcommand{\beq}{\begin{equation}}
\newcommand{\eeq}{\end{equation}}
\newcommand{\bqa}{\begin{eqnarray}}
\newcommand{\eqa}{\end{eqnarray}}

\makeatletter

\usepackage{babel}

\begin{document}
	
	\title{Randomness Certification from Multipartite Quantum Steering for Arbitrary Dimensional Systems}
	\author{Yi Li}
	\address{State Key Laboratory for Mesoscopic Physics, School of Physics, Frontiers Science Center for Nano-optoelectronics, Peking University, Beijing 100871, China}
 \address{Beijing Academy of Quantum Information Sciences, Beijing 100193, China}
	\author{Yu~Xiang}
	\email{xiangy.phy@pku.edu.cn}
	\address{State Key Laboratory for Mesoscopic Physics, School of Physics, Frontiers Science Center for Nano-optoelectronics, Peking University, Beijing 100871, China}
	\address{Collaborative Innovation Center of Extreme Optics, Shanxi University, Taiyuan, Shanxi 030006, China}
	\author{Xiao-Dong, Yu}
	\address{Department of Physics, Shandong University, Jinan 250100, China}
	\author{H. Chau Nguyen}
	\address{Naturwissenschaftlich-Technische Fakult\"{a}t, Universit\"{a}t Siegen, Walter-Flex-Stra{\ss}e 3, 57068 Siegen, Germany}
	\author{Otfried G\"uhne}
	\address{Naturwissenschaftlich-Technische Fakult\"{a}t, Universit\"{a}t Siegen, Walter-Flex-Stra{\ss}e 3, 57068 Siegen, Germany}
	\author{Qiongyi~He}
	\address{State Key Laboratory for Mesoscopic Physics, School of Physics, Frontiers Science Center for Nano-optoelectronics, Peking University, Beijing 100871, China}
	\address{Collaborative Innovation Center of Extreme Optics, Shanxi University, Taiyuan, Shanxi 030006, China}
	\address{Peking University Yangtze Delta Institute of Optoelectronics, Nantong 226010, Jiangsu, China}		
	\address{Hefei National Laboratory, Hefei 230088, China}
	
	\begin{abstract}
		Entanglement in bipartite systems has been applied for the generation of secure random numbers, which are playing an important role in cryptography or scientific numerical simulations. Here, we propose to use multipartite entanglement distributed between trusted and untrusted parties for generating randomness of arbitrary dimensional systems. We show that the distributed structure of several parties leads to additional protection against possible attacks by an eavesdropper, resulting in more secure randomness generated than in the corresponding bipartite scenario. Especially, randomness can be certified in the group of untrusted parties, even there is no randomness exists in either of them individually. We prove that the necessary and sufficient resource for quantum randomness in this scenario is multipartite quantum steering when two measurement settings are performed on the untrusted parties. However, the sufficiency no longer holds with more measurement settings. Finally, we apply our analysis to some experimentally realized states and show that more randomness can be extracted in comparison to the existing analysis. 
\end{abstract}
\maketitle
\textit{Introduction.---}Randomness plays an important role in scientific simulation and cryptography~\cite{rmp_2017_random,npj_2016_review_random}. Different from the classical theory, where any system admits at least a deterministic description, measurements in quantum mechanics have an inherently random character~\cite{born_1926}. As another remarkable feature of quantum theory, entanglement can be used to certify randomness. For example, measurement outcomes leading to a Bell inequality violation cannot be deterministically predicted within any no-signaling theory~\cite{bell_1964, rmp_2014_nonlocality,pra_2006_acin}, thus intrinsically randomness exists among the outcomes. Therefore, some protocols for randomness generation were recently derived from this feature~\cite{prl_2022_roger,njp_2015_xiongfengma,pra_2013_bipartite,prl_2015_acin,prr_2020_acin,prl_2018_paul,prl_2012_acin,pra_2013_multipartite,quantum_2018_multipartite,njp_2014_the,jpa_2014_the,njp_2015_paul,pra_2022_the} and demonstrated in experiments~\cite{prl_2019_shumingcheng,science_2018_jianweiwang,pra_2022_exp,PhysRevLett.126.050503,optica_2016_feihuxu,prl_2017_sdi,qst_2017_paul}. 
	
	%%%%%%%%%%%%%%%%%%%%%%%%%%%%%%%%%%%%%%%%%%%%%%%%%%%%%%%%%%%%%%%%5
	\begin{figure}[t]
		\includegraphics[width=0.45\textwidth]{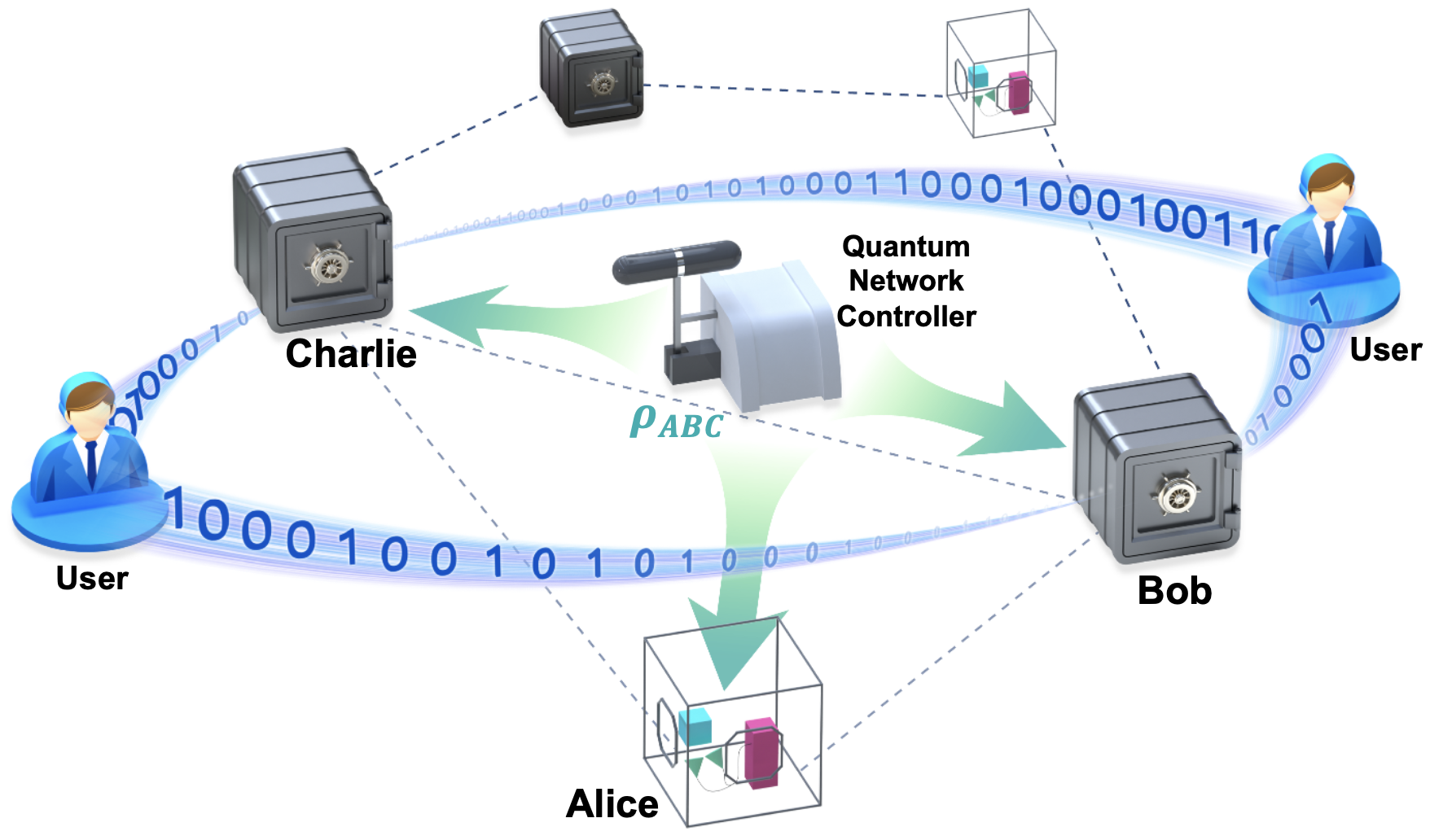}	\caption{\label{fig:scheme}Schematic view on randomness generation in a multipartite network as discussed in this paper. A controller sends a tripartite state $\rho_{ABC}$ to three nodes. Two of these nodes (Bob and Charlie) perform measurements with the aim to use the results as a source of randomness. The measurements of Bob and Charlie are not characterized, consequently they are represented by black boxes. A third trusted party (Alice) performs well-characterized measurements, determining the set of conditional states, thereby limiting the potential attacks by an eavesdropper. We find that the separation between Bob and Charlie allows to generate more randomness than as if they are grouped together; the separation limits the observed correlations between them, but the potential attacks of an eavesdropper become even more limited. If Bob and Charlie have two measurement settings only, then randomness generation is equivalent to quantum steering.}
	\end{figure}

Quantum steering is an intermediate type of quantum correlations between inseparability~\cite{rmp_2009_entanglement} and Bell nonlocality~\cite{bell_1964}. It describes the phenomenon that one party can remotely adjust the states of the other if they are entangled~\cite{rmp_2020_steering,rpp_2016_paul,prxuantum_2022_yuxiang}. In such a scenario, the entanglement can be verified without relying on any assumed models of the steering party’s devices~\cite{prl_2007_wiseman}. This leads to a one-sided device-independent approach to certify randomness~\cite{njp_2015_paul}, which is more robust to noise than the fully-device-independent protocols based on a Bell inequality violation~\cite{nature_2016_review_random,np_2021_ustc,nphy_2021_random,prl_2018_ustc,prl_2018_christian,nature_2018_superluminal,nature_2018_ustc,prl_2020_random,nature_2010_random}.

In view of a potential real-world quantum network distributing multipartite entanglement, it is a relevant topic to explore the generation of randomness distributed over many nodes in an entanglement-based network. So far, multipartite quantum steering~\cite{prl_2013_qiongyihe,nc_2015_multipartite} has been successfully demonstrated in photonic networks~\cite{nc_2015_multipartite,prl_2015_gme,prl_2020_network}, continuous-variable optical networks~\cite{np_2015_cvnetwork,prl_2017_cvnetwork,prl_2020_cvnetwork,prr_2020_cvnetwork}, and atomic ensembles~\cite{science_2018_atom}. The majority of theoretical studies and experiments for randomness generation, however, have focused specifically on the bipartite  scenario~\cite{njp_2015_paul,science_2018_jianweiwang,pra_2022_exp,pra_2022_the,prl_2019_shumingcheng}, where a well-known theorem by Schrödinger  \cite{mpcps_1936_schro,pr_1957_ghjw,pla_1993_ghjw,hpa_1989_ghjw} guarantees that any no-signalling state assemblage can originate from a global quantum state. Consequently, the considered task can be expressed in terms of a semidefinite programming (SDP) problem over all no-signaling bipartite assemblages. This approach, 
	however, cannot be extended to the multipartite case, since the aforementioned equivalence ceases to hold~\cite{prl_2017_postquantum}.
	
Moreover, in order to determine the minimal resources required for quantum cryptography, and also for fundamental interest, the relationship between quantum correlations and randomness has been discussed. Great effort has been devoted to demonstrating that entanglement, steering and nonlocality are {\it necessary} for certifying randomness~\cite{prl_2018_paul,njp_2014_the,prl_2019_shumingcheng,pra_2022_the,quantum_2018_multipartite,pra_2013_multipartite,jpa_2014_the,prl_2012_acin,njp_2015_paul}, but the {\it quantitative} connections are subtle ~\cite{prl_2012_acin,prl_2015_acin,prl_2018_paul,prr_2020_acin}. In these cases, the untrusted parties implemented two measurements only, but increasing the number of measurement settings could bring many benefits, e.g., additional nonlocal and steerable states can be found~\cite{prl_2010_belltest,prl_2022_belltest,prl_2022_shareability,pra_2013_monogamy,pra_2008_Vertesi,np_2010_howard}. For the general cases, however, whether nonlocality or steering is \textit{sufficient} for certifying randomness remains elusive.
	
In this paper, we present the certification of randomness in multipartite quantum systems of any dimension. As shown in Fig.~\ref{fig:scheme}, the scenario we consider is close to the actual situation where only a few of the users have knowledge of their measurement apparatuses (transparent boxes) while the remaining users do not (black boxes). Qualitatively, from the definition of multipartite steering~\cite{nc_2015_multipartite}, we prove that multipartite steering with two-setting measurements on the untrusted nodes is \textit{necessary and sufficient} for certifying randomness in the asymmetric network, independent of the number of outcomes and parties. In the case of more than two settings, this perfect equivalence is broken; some states become steerable but cannot be used to certify randomness. As mentioned above, directly quantifying the amount of randomness cannot be equivalent to an SDP problem for the multipartite scenario. So we calculate lower bounds for the multipartite certified randomness in discrete-variable and continuous-variable systems using the Navascu\'{e}s-Pironio-Ac\'{\i}n (NPA) hierarchy~\cite{prl_2007_npa,njp_2008_npa}, which tests for membership in the set of quantum behaviors. In order to demonstrate the tightness of the lower bounds, upper bounds are calculated by fixing the dimension of each system. We show that certain scenarios, each individual party cannot have certified randomness but, surprisingly, eavesdroppers cannot attack them simultaneously. That means, they can still collaborate to generate joint secured randomness.	Finally, we adopt some existing experimental data~\cite{prl_2022_shareability} to certify randomness, which show that more randomness can be generated with our multipartite scenario than previous experiments in the bipartition scenario~\cite{qst_2017_paul}.

%%%%%%%%%%%%%%%%%%%%%%%%%%%%%%%%%%%%%%%%%%%%%%%%%5
\textit{Randomness in multipartite quantum networks.---}We focus on a tripartite scenario, in which three parties, Alice, Bob and Charlie, are located in distant laboratories and receive an unknown tripartite entangled state $\rho_{ABC}$ from the Controller, as shown in Fig.~\ref{fig:scheme}. Neither Bob nor Charlie trusts their devices, which are consequently treated as ``black boxes''. Still, their measurements are given by an unknown positive operator valued measure (POVM), which is a set of positive semi-define matrices $\{M_i\}_i$ that satisfies $\sum_i M_i = I$. Bob and Charlie apply measurements $M_{b|y}$ and $M_{c|z}$ labeled by $y\in\{0,\cdots,m_B-1\}$ and $z\in\left\{0,\cdots,m_C-1\right\}$, then generate outputs $b\in\left\{0,\cdots,n_B-1\right\}$ and $c\in\left\{0,\cdots,n_C-1\right\}$, respectively. 
	The third party, Alice, has complete knowledge of her device, which allows her to perform quantum state tomography, and thus to obtain a set of unnormalized states $\sigma_{bc|yz} = {\rm Tr}_{ BC}\left[I_A \otimes M_{b|y} \otimes M_{c|z} \rho_{ABC}\right]$ (referred to as a state assemblage) conditioned on Bob's and Charlie's measurements and results.
	
We assume a potential eavesdropper, Eve, who has access to her part of a quadripartite state $\rho_{ABCE}$ and wants to predict the outcomes $b$ and $c$ simultaneously, while giving the measurement  choices $y^*$ and $z^*$ for Bob and Charlie. Since Eve knows which measurements Bob and Charlie will choose to extract randomness, she can optimize her attack to obtain information about these outcomes but still needs to be in line with the observed assemblage. Consequently, Eve gives guesses $e \in \left\{ 0, 1, \cdots, n_B - 1\right\}$ and $e^\prime \in \left\{ 0, 1, \cdots, n_C - 1\right\}$ by performing a POVM measurement $\{M_{e,e'}\}_{e,e'}$. The total guessing probability that Eve's guesses $e=b$ and $ e^\prime=c $ is given by
	\begin{equation}
	\label{eq:1}
	P_{g}(y^*,z^*) = \sum_{e, e^\prime} P_{BCE}(b = e,c = e^\prime,e,e^\prime|y^*,z^*).
	\end{equation}
	Hence randomness, quantified by the min-entropy~\cite{Ieee_2009_Hmin} $H_{\min}=-\log_2(P_g(y^*,z^*))$, can be certified whenever the guessing probability $P_g<1$. This means Eve cannot be completely sure of both Bob's and Charlie's measurement results simultaneously.
	
In order to figure out the optimal strategy for Eve, we maximize her guessing probability (\ref{eq:1}) over all measurement strategies and the possible state accessible to her, which results in the following optimization problem: 
	\begin{align}
	\label{eq:2}
	&\max~~~~P_{g}(y^*,z^*) = \sum_{e, e^\prime}{\rm Tr}\left[  \left(  M_{b = e|y^*} \otimes M_{c = e^\prime|z^*} \otimes M_{e ,e^\prime} \right)  \rho_{BCE} \right] \nonumber\\
	&{\rm w.r.t.} ~~~~ \rho_{ ABCE},~\{M_{b|y}\}_{b,y},~\{M_{c|z}\}_{c,z},~ \{M_{e,e^\prime}\}_{e ,e^\prime}\nonumber\\
	&\operatorname{ s.t.} ~~~~\operatorname{Tr}_{BC} \left[\left(I_A\otimes M_{b |y} \otimes M_{c |z}\right) \rho_{ABC} \right] = \sigma_{b c|y z}^{obs}, ~~\forall b,c,y,z, \nonumber\\
	&\qquad ~~~~\rho_{ABCE} \geq 0 , \quad \operatorname{Tr}\left[\rho_{ ABCE}\right] = 1,\nonumber\\
	&\qquad ~~~~\left\{M_{b|y}\right\}_{b},~ \left\{M_{c|z}\right\}_{c},~\left\{M_{e,e^\prime}\right\}_{e, e^\prime}\in {\rm POVM}, \quad \forall y,z,
	\end{align}
	where $\rho_{BCE}={\rm Tr}_A[\rho_{ABCE}]$, $\rho_{ABC} = {\rm Tr}_{E}[\rho_{ABCE}]$ and $\{\sigma_{bc|yz}^{obs}\}_{b,c,y,z}$ is the assemblage observed by Alice. Note that the first constraint guarantees that the entire state is compatible with the assemblage observed by Alice.   %\cn{I move this sentence down; still in the comments in latex here} %However, this problem is difficult to solve, since the dimension of Bob, Charlie and Eve are unknown.
	
\textit{Multipartite steering as a resource for certified randomness.---}Multipartite steering is defined when both Bob and Charlie hold the untrusted devices and the assemblage $\{\sigma_{bc|yz}^{obs}\}_{b,c,y,z}$ cannot be explained by a fully separable model, i.e., $\rho^{{A}: {B}: {C}}\neq\sum_\lambda p_\lambda \rho_\lambda^{{A}} \otimes \rho_\lambda^{{B}} \otimes \rho_\lambda^{{C}}$. For this, strong tests in terms of SDPs exist \cite{nc_2015_multipartite,rpp_2016_paul}. Combining this definition as well as certifiable randomness, we find that multipartite steering is \textit{necessary} for the certification of multipartite randomness on Bob and Charlie. Specifically, in the case of $m_B = m_C = 2$, multipartite steering is \textit{necessary and sufficient} for certifying randomness. However, in the case of more settings, sufficiency no longer holds. 
	
The ideas of the arguments are as follows: (1) Since the assemblage $\sigma_{bc|yz}^{obs}$ is unsteerable if it can be described by a local hidden state model, where the distribution can be written as a convex sum of local deterministic distributions~\cite{rpp_2016_paul}, the existence of multipartite steering is a necessary condition for generating randomness.
(2) For the reverse direction, we start with the case of $m_B = m_C = 2$, i.e., Bob measures $\{y^*, \bar{y}^*\}$ and Charlie measures $\{z^*,\bar z^*\}$. No verifiable randomness on Bob and Charlie’s sides means that Eve can predict their outcomes of measurements $y^*$ and $z^*$ perfectly, which implies the conditional states at Alice's side generated by $y^*$ and $z^*$ is the same as that generated by Eve’s measurement $M_{e,e^{\prime}}$. Thus, the state assemblage observed by Alice can be seen as generated from the set of measurements $\{M_{e,e^{\prime}}, \bar y^*, \bar z^*\}$.
Eve’s measurement $M_{e,e^{\prime}}$ is, however, compatible with Bob's and Charlie's  measurements $\bar y^*$ and $\bar z^*$ since they are made locally on separate parties. This compatibility ensures that the joint probability distribution of Bob and Charlie is local~\cite{prl_2014_brunner,prl_2014_otfried}, and thus the assemblage is unsteerable by independent Bob and Charlie~\cite{prl_2015_otfried,rmp_2023_otfried}. Hence, the fact that quantum steering in an actual multipartite scenario is sufficient to certify nonzero randomness is proved; more details can be found in Appendix~A. %Supplemental Material~\cite{supp}. 
However, the proof also shows that this sufficiency can be broken with more settings. For instance, when $m_B\geq3$, the additional measurement settings can bring incompatibility to the set of Bob's measurements (expressing  steerability~\cite{prl_2014_brunner,prl_2014_otfried}). But it doesn't affect Eve's unit guessing probability for $y^*$ (still zero randomness). Notice that the above argument is generally valid for multipartite as well as bipartite scenarios.  
	
\textit{Quantification of certified randomness in multipartite scenarios.---} Steering-based randomness in multipartite scenarios was first studied in Ref.~\cite{qst_2017_paul} by considering bipartitions of the W state, in which the measurements performed by Bob and Charlie are global, i.e., $M_{(bc)|(yz)}\neq M_{b|y}\otimes M_{c|z}$. This can be considered as a special case of bipartite scenario, where the task of randomness certification can be expressed in terms of an SDP over all no-signaling bipartite assemblages within quantum theory~\cite{njp_2015_paul}. However, in an actual tripartite scenario where the measurements performed by Bob and Charlie are local, there exist assemblages $\{\sigma_{bc|yz}\}_{b,c,y,z}$ that satisfy the no-signaling principle but do not admit a quantum realization~\cite{prl_2017_postquantum}. The technique to reduce problem~\eqref{eq:2} to an SDP thus generally fails for the multipartite scenario.
In the following, we introduce a simplification of the problem (\ref{eq:2}), which subsequently allows for derivation of upper and lower bounds based on a see-saw application of SDPs and the NPA hierarchy, respectively (see Appendix~B).%~\cite{supp}.
	
Firstly, since Eve only implements a single POVM, we can always use a joint classical-quantum state~\cite{rmp_2017_entropy} $\rho_{ABCE} = \sum_{e,e^\prime} |e,e^\prime\rangle_E \langle e,e^\prime| \otimes \sigma_{ABC}^{e e^\prime}$ to describe the behavior of the partites without loss of generality. Here, $\sigma_{ABC}^{ee^\prime}$ is an unnormalized quantum state conditioned on Eve's outcome $e,e^\prime$. Thus, the maximization problem (\ref{eq:2}) can be simplified to maximize $\sum_{e, e^\prime}\operatorname{Tr}[  ( I_A \otimes M_{b = e|y^*} \otimes M_{c = e^\prime|z^*} )  \sigma_{ABC}^{ee^\prime} ]$ by searching for the triple $\{\sigma_{ABC}^{ee^{\prime}}, M_{b|y}, M_{c|z}\}$, where the dimension of Eve's system is not relevant anymore. 
	
Then, upper bounds on the randomness $H_{\min}^{Dim}$ with fixed dimension can be achieved by optimizing over individual variables of the triple, each corresponding to a SDP (see-saw algorithm). Furthermore, a lower bound $H_{\min}^{NS}$ can be obtained by relaxing the constraints on Eve to the impossibility of superluminal signaling. This means that $H_{\min}^{NS}$ can be calculated by solving an SDP problem over an assemblage $\sigma_{bc|yz}^{ee'} = {\rm Tr}[I_A\otimes M_{b|y}\otimes M_{c|z}\sigma_{ABC}^{ee'}]$ with no-signaling constraint.
	
Finally, in order to give a more realistic range of quantum realizations, the generalized NPA hierarchy~\cite{rpp_2016_paul,prl_2017_postquantum} provides a series of tests which an assemblage must pass if it admits a quantum realization. Hence some lower bounds $H_{\min}^{Q_k}$ can be calculated by replacing the constraints from no-signaling set to the $Q_k$ sets, where $k$ corresponds to different NPA levels. See more details in Appendix~B. %Supplemental Material~\cite{supp}.
We find that by optimizing $\{M_{b|y}, M_{c|z}\}$ independently, an actual multipartite scenario can bring more randomness than the bipartition scenario $H_{\min}^{Glo}$ with optimizing global measurement $\{M_{(bc)|(yz)}\}$, although they are both multiple parties involved.
	
Besides, in the multipartite scenario, the amount of randomness generated on either party can also be considered individually. Now Eve only guesses the measurement outcomes on one of the untrusted parties. Therefore, the randomness solely on Bob's outcomes can be certified by changing the objective function of Eq.~(\ref{eq:2}) into $\sum_{e}\operatorname{Tr}[(  M_{b = e|y^*} \otimes M_{e} )  \rho_{BE}]$, where $\rho_{BE} = {\rm Tr}_{ AC}[\rho_{ABCE}]$, and so for Charlie. Note that this randomness is still constrained by the observed assemblage $\{\sigma_{bc|yz}^{obs}\}_{b,c,y,z}$ in a tripartite scenario, which is different from the previous bipartite case. Similarly, we can derive upper and lower bounds for the separate randomness generated only on Bob (or Charlie), more details can be found in Appendix~B.

 \begin{figure}[b]
\includegraphics[width=0.49\textwidth]{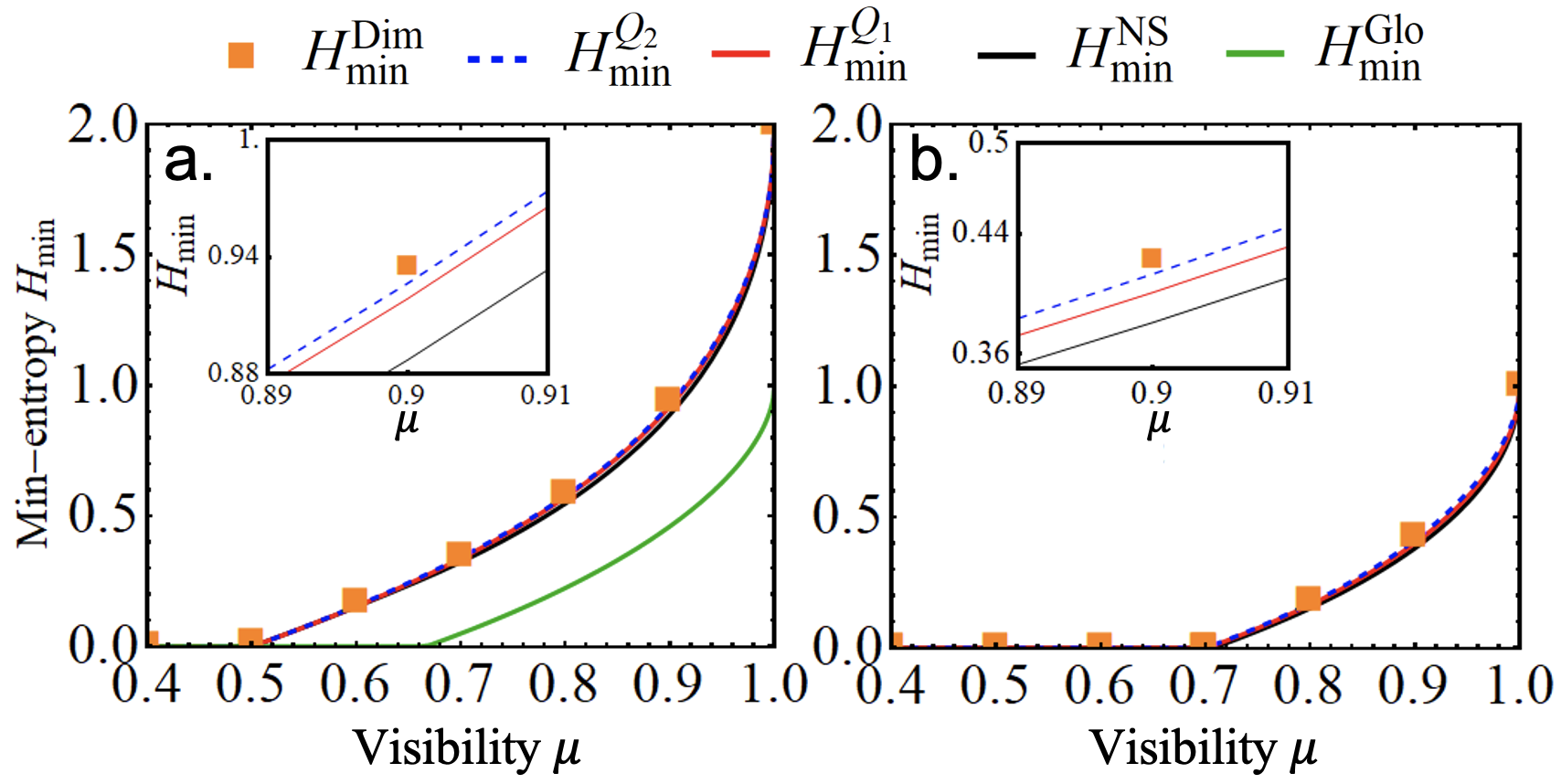}
\caption{\label{fig:GHZ_Result} Multipartite randomness certified on Bob and Charlie together (a) and the separate randomness on Bob or Charlie only (b) in GHZ states with $N=3$, $d=2$. Upper bounds with fixed dimensions $d_B=d_C=10$ (orange square) are closed with the lower bounds that corresponds with different NPA levels (red solid curve for $Q_1$ and blue dashed line for $Q_2$). The lowest bound (black solid curve) is constrained by the no-signaling principle. The difference between the upper bound and the maximum lower bound is about $10^{-3}$, which means the lower bounds are tight~\cite{footnote}. The green solid curve shows the certified randomness when Bob and Charlie's measurements are global.}
\end{figure}

Now, we apply our findings to various experiment-relevant  multipartite states, from discrete-variable to continuous-variable systems.

\paragraph{(i) GHZ state.---}Consider a $d$-dimension GHZ state over $N$ subsystems mixed with white noise, $\rho_{\mu} = \mu|\Psi\rangle\langle\Psi| + \frac{1-\mu}{d^N}\mathbb{I}$,
	where $|\Psi\rangle = {\frac{1} {\sqrt{d}}}\sum_{i=0}^{d-1}|i\rangle^{\otimes N}$ and visibility $\mu\in[0,1]$. 
	Starting with the simplest case of $N=3$ and $d=2$, Bob and Charlie both perform three Pauli measurements $\{\hat{X},\hat{Y},\hat{Z}\}$, and the assemblages $\{\sigma_{bc|yz}^{obs}\}_{b,c,y,z}$ are observed by Alice's tomography.  
	Figure~\ref{fig:GHZ_Result}(a) shows upper and lower bounds for the min-entropy of the certifiable randomness on Bob and Charlie's outcomes generated by $y^* = z^* = \hat{X}$. In particular, the min-entropy is positive for $\mu > 0.5$ and achieves its maximum of 2 bits at $\mu = 1$. Compared with the randomness $H_{\min}^{Glo}$ for the bipartition scenario, more randomness can be certified.

	Figure~\ref{fig:GHZ_Result}(b) shows the separate randomness generated solely on Bob's (or Charlie's) side , which exists in the region of $\mu>0.70$. 
	Compared with Fig.~\ref{fig:GHZ_Result}(a), certifying randomness only in one party requires higher state visibility.
	In particular, when $0.50 < \mu \leq 0.69$, there exists nonzero randomness on the untrusted parties together, even though no separate randomness is induced in either parties individually, which leads to additional protection against possible attacks. 
	
	\begin{figure}[b]
		\includegraphics[width=0.4\textwidth]{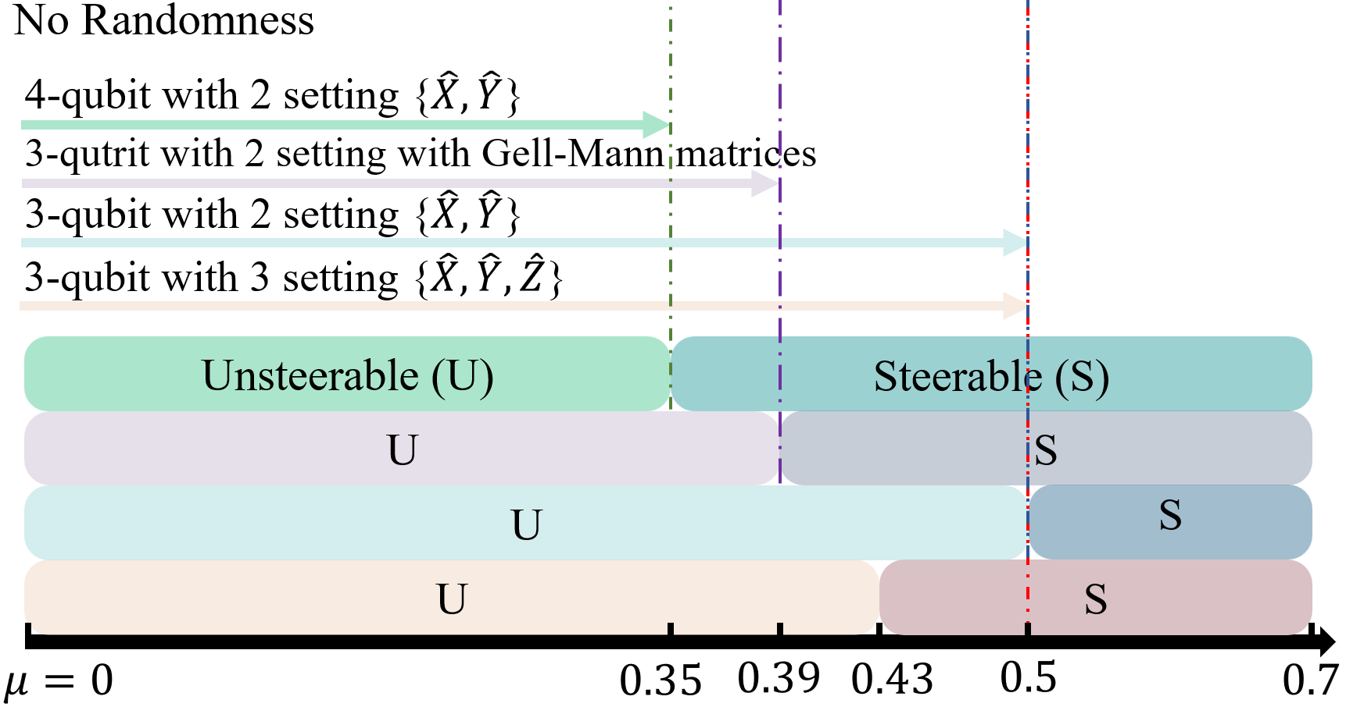}		\caption{\label{fig:bound} The relationship of thresholds between the randomness on Bob and Charlie together and multipartite steering (from Bob and Charlie to Alice) in different GHZ states with two or three measurement settings. The left (right) blocks mean the observed assemblage is unsteerable (steerable). The arrows mean there is no randomness in their corresponding range of visibility. For the two-setting cases, the thresholds of randomness always agree with that of multipartite steering. However, for the three-setting case, the threshold of multipartite steering is decreased to 0.428 while the threshold of randomness remains unchanged.}
	\end{figure}
 
We further investigate general cases of GHZ states with different numbers of parties $N$ and dimensions $d$. For four parties, the measurements of three nodes are not characterized while the well-characterized measurements are performed by the rest node. The results are shown in Fig.~\ref{fig:bound}, which agree with our above qualitative discussions. In particular, it is clearly seen that for the case of two-setting measurements, the thresholds of multipartite randomness (generated on Bob and Charlie together) are consistent with the condition for showing multipartite steering. Observe that increasing the number of measurements decreases the thresholds of multipartite steering for the 3-qubit GHZ state from $0.5$ (with measurements $\hat{X}$ and $\hat{Y}$) to $0.428$ (with measurements $\hat{X}$, $\hat{Y}$, and $\hat{Z}$). However, the threshold for certified randomness remains at $0.5$ even for three-setting measurements. 

	\paragraph{(ii) W-like state with experiment data.---} In Ref.~\cite{prl_2022_shareability}, a class of W-like states $|\Psi_W\rangle = \alpha|001\rangle_{\rm ABC} + \beta|010\rangle_{\rm ABC} + \gamma|100\rangle_{\rm ABC}$ were experimentally implemented to demonstrate the sharability of quantum steering with different measurement settings.  
		Adopting their tomographic data for  $(\alpha,\beta,\gamma) = (0.575,0.582,0.576)$, nearly a W state, we calculate the amount of randomness for different scenarios.
		The results are listed in Table \ref{table:1}. It can be seen that the amount of reliable random bits $H_{\min}^{Q_2}$ certified by local measurements is significantly higher than that by the method with global measurements adopted in the previous experiment~\cite{qst_2017_paul}.%, 
	
	\begin{table}[t]
		\centering
		\renewcommand\arraystretch{1.5}
		\caption{Randomness certified on different parties with the experimental data in Ref. \cite{prl_2022_shareability}. Here the optimal measurements are chosen %$y^* = \hat{X}$ and $z^* = \hat{Y}$ for tripartite scenario and  $y^* = \hat X$ and $z^* = \hat Z$ for.
		to maximize the randomness. }\label{table:1}
		\vspace{2mm}
		\resizebox{\linewidth}{!}{
			\begin{tabular}{| c | c | c | c | c | c |}
				\hline\hline
				Parties of Randomness & $H_{\rm min}^{NS}$ &  $H_{\rm min}^{Q_1}$&  $H_{\rm min}^{Q_2}$&  $H_{\rm min}^{Dim}$ &  $H_{\rm min}^{Glo}$\\ \hhline{*{6}{-}}
				Bob Only &	0.592 &	0.736&	0.738 &0.769 & N/A \\\hhline{|-|-|-|-|-|-|}
				Charlie Only &	0.595  & 0.739 & 0.740 & 0.774 & N/A \\\hhline{|-|-|-|-|-|-|}
				Bob \& Charlie & 1.236 & 1.445 &	1.451 &	1.525 &	0.783 \\\hhline{*{6}{:=}:}
			\end{tabular}
		}
	\end{table}
 
	\begin{figure}[b]
		\includegraphics[width=0.48\textwidth]{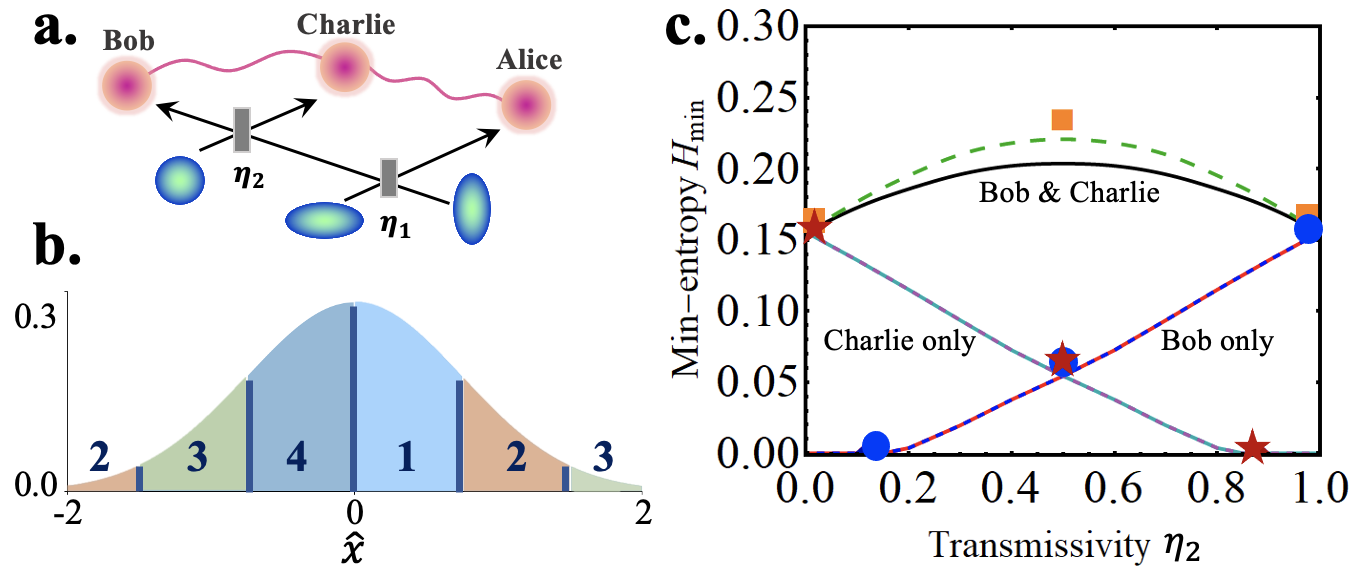}
		\caption{\label{fig:CV_Result} Randomness certified in continuous-variable system. (a) Scheme of generating pure three-mode entangled Gaussian state by a linear optical network. (b) Bob's quadrature measurement $\hat{x}$ is binned into 4 outcomes. (c) The lower (solid curve for $H_{\min}^{NS}$ and dashed curve for $H_{\min}^{Q_1}$) and upper bounds (orange square for Bob and Charlie together, red pentagram for Charlie only and blue circle for Bob only) of randomness. Here we set $r = 0.345$ (corresponding to -3 dB quadrature noise),  $\eta_1 = 1/2$, $y^*=z^*=\hat x$, and cut off the Fock basis to one photon.}
	\end{figure}
	
	\paragraph{(iii) Three-mode squeezed vacuum state.---}
 A three-mode entangled Gaussian state can be generated by mixing two squeezed inputs with squeezing level $r$ and one extra vacuum state as shown in Fig.~\ref{fig:CV_Result}(a).
	For the system with continuous variables, we can bin the homodyne measurement outputs into a finite number of outcomes
	like Fig.~\ref{fig:CV_Result}(b)~\cite{prl_2018_mubcv,pra_2022_the}. 
	By analyzing the assemblage, we evaluate the upper and lower bounds of randomness on Bob's and Charlie's measurement results as well as the separate randomness on Bob or Charlie only. Note that the min-entropy is maximized over binning periods $T_{\hat x, Bob}, T_{\hat x, Charlie}\in [2,10]$ independently; more details in Appendix~C. %Supplemental Material~\cite{supp}.
	
	As Bob and Charlie always steer Alice together with quadrature measurements $\{\hat{x}, \hat{p}\}$, the multipartite randomness on Bob and Charlie exists for any transmission factor $\eta_2$ of second beam splitter, as illustrated in Fig~\ref{fig:CV_Result}(c). However, when $\eta_2$ is in the range $[0,0.11]$ or $[0.89,1]$, Eve can guess the measurement outcomes of Bob or Charlie correctly with unit probability.
	
	\textit{Conclusion.---}
 We first present the certification of randomness generated from multiple untrusted parties in an asymmetric network and discussed the relation between multipartite steering and verifiable randomness. When the untrusted parties perform two-setting measurements locally, we proved that multipartite steering is necessary and sufficient for generating randomness in such an asymmetric network by connecting the randomness with incompatible measurements. Increasing the measurement setting contributes to demonstrating 
 steering but does not necessarily certifies randomness in a larger parameter range, which helps us to determine the minimal resource in quantum cryptography. Furthermore, we quantified multipartite randomness on some typical states from discrete-variable to continuous-variable systems. The results showed that the amount of multipartite randomness is significantly improved, which can promise additional security in quantum network. So far, multipartite steering has been demonstrated in various platforms~\cite{nc_2015_multipartite,prl_2015_gme,prl_2020_network,np_2015_cvnetwork,prl_2017_cvnetwork,prl_2020_cvnetwork,prr_2020_cvnetwork,science_2018_atom}, which lays a favorable foundation for generating multipartite randomness. Our results make a significant advance in an in-depth understanding of quantum randomness as a fundamental resource and provide an important framework for the multipartite quantum network.
	
	\begin{acknowledgments}
		We acknowledge enlightening discussions with Paul Skrzypczyk and experimental data from Kai Sun. This work is supported by the National Natural Science Foundation of China (Grants No.~11975026, No. 12125402, No.~12004011, and No.~12147148), Beijing Natural Science Foundation (Grant No.~Z190005), the Key R\&D Program of Guangdong Province (Grant No. 2018B030329001), and the Innovation Program for Quantum Science and Technology (Grant No. 2021ZD0301500). X.D.Y. acknowledges support by the National Natural Science Foundation of China (Grants No. 12205170 and No. 12174224) and the Shandong Provincial Natural Science Foundation of China (Grant No. ZR2022QA084). H.C.N. and O.G. were supported by the Deutsche Forschungsgemeinschaft (DFG, German Research Foundation, project numbers
447948357 and 440958198), the Sino-German Center for Research Promotion (Project M-0294), the ERC (Consolidator Grant 683107/TempoQ), the German Ministry of Education and Research (Project QuKuK, BMBF Grant No. 16KIS1618K).
	\end{acknowledgments}

%%%%%%%%%%%%%%%%%%%%%%%%%%%%%%%%%%%%%%%%%%%%%%%%%%%%%%%%%%%%%%%%%%%%%%%%%%%%%%%%%%%%%%%%%%%%%%%%%%%%%%%%%%%%%%%
\onecolumngrid

\setcounter{equation}{0}
\renewcommand\theequation{A\arabic{equation}}
\section{Appendix A:~Relation between Steering and Randomness}
In this section, we prove that multipartite steering is \textit{necessary} for the certification of multipartite randomness on the untrusted parties. Specifically, in the case of $m_B = m_C = 2$, multipartite steering is \textit{necessary and sufficient} for certifying randomness. This perfect equivalence, however, ceases to hold with more measurement settings. This argument is generally valid for multipartite as well as bipartite scenarios. Here, we first give a proof in the bipartite scenario.

\subsection{1.~Bipartite scenario}
In the bipartite scenario, Alice and Bob receive a bipartite state $\rho_{AB}$ from Controller while they are located in distant laboratories. The measurement device on Bob is untrusted. He can choose which measurement $y\in\{0,\cdots,m_B-1\}$ to perform, each of which gives an outcome $b\in\{0,\cdots,n_B-1\}$. On the other side, Alice has complete knowledge of her device, which allows her to reconstruct every conditional state $\sigma_{b|y}^{obs} = {\rm Tr}_{B}[ I_A \otimes M_{b|y} \rho_{AB}]$ (unnormalized) she received.

For the observed assemblage, we can detect steering by solving the following SDP problem~\cite{rpp_2016_paul}:
\begin{align}
	\label{eq:S2}
	&{\rm find}\quad  \{\sigma_{\lambda}\}_\lambda\notag\\ 
	&{\rm s.t.} \quad \sigma_{b|y}^{obs} = \sum_{\lambda}D_{Local}(b|y,\lambda)\sigma_{\lambda}, \quad \forall b,y,\\
	&\qquad \sigma_\lambda\geq 0, \quad \forall \lambda.\notag
\end{align}
We can also quantify the amount of randomness on Bob's outputs by solving the SDP problem~\cite{njp_2015_paul}:
\begin{align}
		\label{eq:S3}
		P_{\mathrm{g}}\left(y^*\right)=&\max _{\left\{\sigma_{b \mid y}^e\right\}_{e,b,y}}  \operatorname{Tr} \sum_e \sigma_{b=e \mid y^*}^e \notag \\
		\text { s.t. } &\sum_e \sigma_{b \mid y}^e =\sigma_{b \mid y}^{\mathrm{obs}}, \quad \forall b, y, \\
		& \sum_b \sigma_{b \mid y}^e=\sum_b \sigma_{b \mid y^{\prime}}^e, \quad \forall e, y, y^{\prime}, \notag\\
		& \sigma_{b \mid y}^e \geqslant 0, \quad \forall b, y, e .\notag
\end{align}
Next, we will show that if problem Eq.~\eqref{eq:S2} is feasible, then problem Eq.~\eqref{eq:S3} gives $P_g(y^*)=1$ for any measurement settings $m_A$. Specifically, in the case of $m_A=2$, problem Eq.~\eqref{eq:S2} must be feasible if problem Eq.~\eqref{eq:S3} gives $P_g(y^*)=1$.

\textit{Steering is necessary for randomness.-} If the problem Eq.~\eqref{eq:S2} is feasible, we have
\begin{gather}
\begin{split}
		\sigma_{b|y}^{obs} = \sum_{\lambda}D_{Local}(b|y,\lambda)\sigma_{\lambda} = \sum_{i=0}^{n_B-1}\sum_{\lambda\in\lambda^{(i)}}D_{Local}(b|y,\lambda)\sigma_{\lambda}, \quad \forall b,y,
\end{split}
\end{gather}
where the deterministic probability distribution is $D_{Local}(b|y,\lambda) = \delta_{b,\lambda_{(y)}}$ and $\lambda = (b_0,b_1,\cdots,b_{m_B-1})$. We can divide all the extremal points of the local set into $n_B$ classes: $\lambda^{(i)}:=\{\lambda|\lambda_{(y^*)} = i\}$ with $ i=0,1,\cdots n_B-1$, and construct an ensemble $\sigma_{b|y}^e = \sum_{\lambda\in\lambda^{(e)}}D_{Local}(b|y,\lambda)\sigma_{\lambda}$, $\forall b,e,y$. Then, we can easily check that such an ensemble gives $P_g(y^*)=1$ in Problem Eq.~\eqref{eq:S3}:
\begin{equation}
	\begin{aligned}
         P_g(y^*) &= {\rm Tr}\sum_{e}\sigma_{e|y^*}^e=\sum_{\lambda}p(\lambda) = 1,\\
		\sum_e\sigma_{b|y}^e & = \sum_{\lambda}D_{Local}(b|y,\lambda)\sigma_{\lambda} = \sigma_{b|y}^{obs}, \quad \forall b,y,\\
		\sum_b\sigma_{b|y}^e &= \sum_{\lambda\in\lambda^{(e)}}\sigma_{\lambda} = \sum_{b}\sigma_{b|y'}^e,\quad \forall e,y,y'.
	\end{aligned}
 \end{equation}
Hence, the fact that no randomness on Bob's outputs if the assemblage measured by Alice is unsteerable for any measurement settings $m_B$ is proved.

\textit{Steering is sufficient for randomness.-} In the case of $m_B = 2$, we set that $y\in\{y^*,\bar{y}^*\}$. If problem Eq.~\eqref{eq:S3} gives
	\begin{align}
		P_g(y^*) = {\rm Tr}\sum_{e}\sigma_{e|y^*}^e = \sum_e p(e,e|y^*) = \sum_e p(e|y^*,e) p(e) = 1,
	\end{align}
where $p(e,e|y) = p(e|y,e)p(e|y) = p(e|y,e)p(e)$ due to the no-signaling condition between Eve and Bob. Then we have $p(b|y^*,e)=\delta_{b,e}$, $\forall e,b$ without loss of generality, which means $\sigma_{b|y^*}^e = p(e)p(b|y^*,e)\rho_{b|y^*}^e=\delta_{b,e}\sigma_{b|y^*}^e$, $\forall e,b$. Further, since the optimal solution always satisfies with the observed assemblage and also the no-signaling condition, then 
\begin{equation}
\begin{aligned}
		\sigma_{b|y^*}^{obs} &= \sum_e \sigma_{b|y^*}^e = \sum_l\sigma_{l|\bar{y}^*}^b = \sum_{q}\sigma_{\lambda_{(y^*,\bar y^*)} = (b,q)} = \sum_{\lambda} D_{Local}(b|y^*,\lambda)\sigma_{\lambda}  , \quad \forall b, \\  \sigma_{b|\bar{y}^*}^{obs} &= \sum_e\sigma_{b|\bar{y}^*}^e = \sum_{q} \sigma_{\lambda_{(y^*,\bar y^*)} = (q,b)} = \sum_{\lambda} D_{Local}( b|\bar{y}^*,\lambda)\sigma_{\lambda} , \quad \forall b.
  \end{aligned}
\end{equation}
where $\sigma_{\lambda_{(y^*,\bar y^*)} = (b,q)} = \sigma_{q|\overline{y}^*}^b\geq 0$. Therefore, in the two-setting measurement case, if there is no randomness on Bob’s outputs, then $\{\sigma_{b|y}^{obs}\}_{b,y}$ is unsteerable.

In fact, $P_g(y^*) = 1$ gives $\sigma_{b|y^*}^e = {\rm Tr}_{EB}[M_e\otimes I_A\otimes M_{b|y^*} \rho_{EAB}] = \delta_{e,b} {\rm Tr}_{EB}[M_e\otimes I_A \otimes M_{b|y^*} \rho_{EAB}]$, $\forall b, e$. Then
	\begin{align}
		\sum_l {\rm Tr}_{EB}[M_b\otimes I_A\otimes M_{l|y^*} \rho_{EAB}]= \sum_{l} {\rm Tr}_{EB}[M_l\otimes I_B \otimes M_{b|y^*} \rho_{EAB}] = \sigma_{b|y^*}^{obs} = \sigma_{b|y^*}^{b},  \quad \forall b,
	\end{align}
which means the measurement $y^*$ on Bob can be regarded as a measurement performed on Eve: $\sigma_{b|y^*}^{obs} = {\rm Tr}_{EB}[M_b\otimes I_A\otimes I_B \rho_{EAB}] = {\rm Tr}_{EB}[I_E\otimes I_A\otimes M_{b|y^*} \rho_{EAB}]$. For the other measurement $\bar{y}^*$, $\sigma_{b|\bar{y}^*}^{obs} = {\rm Tr}_{EB}[I_E\otimes I_A \otimes M_{b|\bar{y}^*}\rho_{EAB}]$. Therefore, the assemblage must be unsteerable since measurements $\{M_e\otimes I_B\}_{e}$ and $\{I_E\otimes M_{b|\bar{y}^*}\}_{b}$ are compatible~\cite{prl_2014_otfried,prl_2014_brunner,prl_2015_otfried}. However, in the case of $m_B\geq 3$, the additional measurements $\{M_{b|\bar{y}^{*'}}\}_b$ could be incompatible with measurement $\bar{y}^*$ and hence express steerability, but do not involve in generating the randomness, i.e. Eve still gives $P_g(y^*)=1$.

\subsection{2.~Tripartite scenario}
Now we generalize the above proofs to the tripartite scenario, we will show if the assemblage has no multipartite steering, problem Eq.~(2) in the main text would give $P_g(y^*,z^*) = 1$. Also the opposite direction is true when $m_B=m_C=2$.

\textit{Steering is necessary for randomness.-} Similarly, for the observed assemblage, we can also detect multipartite steering by solving this SDP problem~\cite{nc_2015_multipartite}:
\begin{equation}
    \begin{aligned}
    \label{eq:s8}
        {\rm find}& \qquad \{\sigma_{\mu\nu}\}_{\mu\nu}\\
        {\rm s.t.}& \qquad \sum_{\mu,\lambda}D(b|y,\mu)D(c|z,\nu)\sigma_{\mu\nu} = \sigma_{bc|yz}^{obs}, \qquad \forall b,c,y,z,\\
        &\qquad \sigma_{\mu\nu} \geq 0, \qquad \forall \mu,\nu,
    \end{aligned}
\end{equation}
If problem Eq.~\eqref{eq:s8} is feasible, we have
\begin{equation}
	\begin{aligned}
		\sigma_{bc|yz}^{obs} = \sum_{\lambda}D_{Local}(bc|yz,\lambda)\sigma_{\lambda} = \sum_{e,e'}\sum_{\lambda\in\lambda^{(e,e^\prime)}}D_{Local}(bc|yz,\lambda)\sigma_{\lambda}, \quad \forall b,c,y,z.
	\end{aligned}
\end{equation}
Here the deterministic probability distribution is
\begin{equation}
	\begin{aligned}
		D_{Local}(bc|yz,\lambda)= \begin{cases}1, & \text { if } b=b_y \text { and } c=c_z, \\ 0, & \text { otherwise, }\end{cases}
	\end{aligned}
\end{equation}
where $\lambda=(b_0,b_1,\cdots,b_{m_B-1};c_0,c_1,\cdots,c_{m_C - 1})$ and $\lambda^{(e,e^\prime)}:=\{\lambda|\lambda_{({y^*};{z^*})}=(e;e')\}$, $\forall e,e'$. Similarly, we construct an ensemble $\sigma_{bc|yz}^{ee^\prime} = \sum_{\lambda\in\lambda^{(e,e^\prime)}}D_{Local}(bc|yz,\lambda)\sigma_{\lambda}$. It can be easily checked that such an ensemble gives $P_g(y^*,z^*)=1$ in the problem Eq.~(2) with simplification (Eq.~\eqref{suppeq:18}) in the main text:
\begin{equation}
	\begin{aligned}
         P_g(y^*,z^*) &= {\rm Tr}\sum_{e,e^\prime}\sigma_{ee^\prime|y^*z^*}^{ee^\prime} = \sum_{e,e^\prime}\sum_{\lambda\in\lambda^{(e,e^\prime)}}D_{Local}(ee^\prime|y^*z^*,\lambda)p(\lambda) = \sum_{\lambda}p(\lambda) = 1,\\
		\sum_{e,e^\prime}\sigma_{bc|yz}^{ee^\prime} &= \sum_{e,e^\prime}\sum_{\lambda\in\lambda^{(e,e^\prime)}}D_{Local}(bc|yz,\lambda)\sigma_{\lambda} = \sum_{\lambda}D_{Local}(bc|yz,\lambda)\sigma_{\lambda} = \sigma_{bc|yz}^{obs}, \quad \forall b,c,y,z,\\
		\sigma_{bc|yz}^{ee^\prime} &= \sum_{\lambda\in\lambda^{(e, e^\prime)}}D_{Local}(bc|yz,\lambda)\sigma_{\lambda} = {\rm Tr}_{BC}[I_A\otimes M_{b|y}\otimes M_{c|z} \sigma_{ABC}^{ee'}],\quad \forall b,c,y,z,e,e',
	\end{aligned}
\end{equation}
where
\begin{equation}
	\begin{aligned}
		M_{b|y} &= I_{0}\otimes\cdots\otimes|b\rangle_y \langle b| \otimes \cdots\otimes I_{m_B-1}, \quad \sum_{b}|b\rangle\langle b| = I,\\
		M_{c|z} &= I_0\otimes\cdots\otimes|c\rangle_z \langle c| \otimes \cdots\otimes I_{m_C - 1}, \quad \sum_{c}|c\rangle\langle c| = I,\\
		\sigma_{ABC}^{ee'} &= \sum_{\lambda \in \lambda^{(e,e^\prime)}} \rho_{BC}^{\lambda}\otimes \sigma_{\lambda},\\ 
		\rho_{BC}^{\lambda = (b_0,\cdots,b_{m_B - 1};c_0,\cdots,c_{m_C - 1})} &= |b_0\rangle_{B_0}\langle b_0|\otimes \cdots\otimes|b_{m_B - 1}\rangle_{B_{m_B - 1}}\langle b_{m_B - 1}|\otimes |c_0\rangle_{C_0}\langle c_0|\otimes \cdots\otimes|c_{m_C - 1}\rangle_{C_{m_C - 1}}\langle c_{m_C - 1}|.
	\end{aligned}
\end{equation}
Therefore, multipartite steering is necessary for certifying randomness on Bob and Charlie.

\textit{Steering is sufficient for randomness.-} In the case of $m_B = m_C = 2$, Bob measures $y\in\{y^*,\bar{y}^*\}$ and Charlie measures $z\in\{z^*,\bar{z}^*\}$, respectively. If problem Eq.~\eqref{eq:S3} gives
\begin{equation}
	\begin{aligned}
		P_g(y^*,z^*) = {\rm Tr}\sum_{e,e'}\sigma_{ee'|y^*z^*}^{ee'} = \sum_{e,e'} p(ee',ee'|y^*z^*) = \sum_{e,e'} p(ee'|y^*z^*,ee') p(ee') = 1,
	\end{aligned}
\end{equation}
where $p(ee',ee^\prime|yz) = p(ee'|yz,ee^\prime)p(ee^\prime|yz) = p(ee'|yz,ee^\prime)p(ee^\prime)$ due to no-signaling condition, we have $p(bc|y^*z^*,ee^\prime)  = \delta_{b,e}\delta_{c,e'}$, $\forall e,e^\prime,b,c$. Then, $p(bc|\bar{y}^*{z}^*,ee') = \delta_{c,e'} p(bc|\bar{y}^*{z}^*,ee')$ and $p(bc|y^*\bar{z}^*,ee') = \delta_{b,e}p(bc|y^*\bar{z}^*,ee')$ can also be derived due to the no-signaling condition. Since the optimal solution satisfies with the observed assemblage and also the no-signaling condition, then 
	\begin{align}
		\sigma_{bc|y^*z^*}^{obs} &= \sum_{e,e'} \sigma_{bc|y^*z^*}^{ee^\prime} =  \sigma_{bc|y^*z^*}^{bc} = \sum_{k,l} \sigma_{kl|\bar{y}^*\bar{z}^*}^{bc}, \quad \forall b,c, \notag\\
		\sigma_{bc|\bar{y}^*{z}^*}^{obs} &= \sum_{e,e^\prime}\sigma_{bc|\bar{y}^*{z}^*}^{ee^\prime} =\sum_{e}\sigma_{bc|\bar{y}^*{z}^*}^{ec} = \sum_{k,l} \sigma_{bl|\bar{y}^*\bar{z}^*}^{kc}, \quad \forall b,c,  \\ 
		\sigma_{bc|{y}^*\bar{z}^*}^{obs}&= \sum_{e,e^\prime}\sigma_{bc|{y}^*\bar{z}^*}^{ee^\prime} =\sum_{e^\prime}\sigma_{bc|{y}^*\bar{z}^*}^{be^\prime} = \sum_{k,l}\sigma_{kc|\bar{y}^*\bar{z}^*}^{bl}, \quad \forall b,c, \notag \\ 
		\sigma_{bc|\bar{y}^*\bar{z}^*}^{obs} &= \sum_{e,e^\prime}\sigma_{bc|\bar{y}^*\bar{z}^*}^{ee^\prime} = \sum_{k,l} \sigma_{bc|\bar{y}^*\bar{z}^*}^{kl}, \quad \forall b,c,\notag
	\end{align}
where $\sigma_{b_{\overline{y}^*},c_{\overline{z}^*}|\overline{y}^*\overline{z}^*}^{b_{{y}^*}c_{{z}^*}} = \sigma_{\lambda = (b_{y^*},b_{\overline{y}^*};c_{z^*},c_{\overline{z}^*})}$, i.e. the observed assemblage can be decomposed as $\sigma_{bc|yz}^{obs} = \sum_{\lambda}D_{Local}(bc|yz,\lambda)\sigma_{\lambda}$. Therefore, in the case of $m_B=2$ and $m_C=2$, if there is no randomness can be certified on Bob’s and Charlie's outputs, then $\{\sigma_{bc|yz}^{obs}\}_{b,c,y,z}$ is unsteerable.

The ensemble we concerned is $\sigma_{BCA}^{ee'} = {\rm Tr}_{E}[M_{e,e'}\otimes I_B \otimes I_C \otimes I_A \rho_{EBCA}]$, which can also be obtained by another global state $\rho_{EBCA}' = \sum_{e,e'} |e\rangle_{E_1}\langle e| \otimes |e'\rangle_{E_2}\langle e'| \otimes \sigma_{BCA}^{ee'}$ as well as Eve's measurement $\{|e\rangle_{E_1}\langle e| \otimes |e'\rangle_{E_2}\langle e'|\}_{e,e'}$. Therefore, we can regard Eve as two local parts ($E_1$ and $E_2$) without loss of generality, then $P_g(y^*,z^*) = 1$ gives
\begin{equation}
	\begin{aligned}
		{\rm Tr}[M_{e}^1\otimes M_{e'}^2\otimes M_{b|y} \otimes M_{c|z}\otimes I_A \rho_{E_1E_2BCA}] = \begin{cases} \delta_{e,b}\delta_{e',c}p(b,c,e,e'|y,z), & \text { for } y=y^* \text { and } z=z^*, \\ 
		\delta_{e,b}p(b,c,e,e'|y,z), &\text { for } y=y^* \text { and } z=\bar{z}^*,\\ 
		\delta_{e',c}p(b,c,e,e'|y,z), &\text { for } y=\bar{y}^* \text { and } z={z}^*,\\ 
		p(b,c,e,e'|y,z), & \text { otherwise, }\end{cases} \qquad \forall e,e',b,c,
	\end{aligned}
\end{equation}
which means
\begin{equation}
	\begin{aligned}
		\sum_{k,l} {\rm Tr}_{E_1E_2BC}[M_{b}^1\otimes M_{c}^2\otimes M_{k|y^*} \otimes M_{l|z^*}\otimes I_A \rho_{E_1E_2BCA}] = \sum_{k,l}{\rm Tr}_{E_1E_2BC}[M_k^1\otimes M_l^2\otimes M_{b|y^*} \otimes M_{c|z^*}\otimes I_A \rho_{E_1E_2BCA}],  \quad \forall b,c,\\
		\sum_{k} {\rm Tr}_{E_1E_2BC}[M_{b}^1\otimes M_{e'}^2\otimes M_{k|y^*} \otimes M_{c|\bar{z}^*}\otimes I_A \rho_{E_1E_2BCA}] = \sum_{k}{\rm Tr}_{E_1E_2BC}[M_k^1\otimes M_{e'}^2\otimes M_{b|y^*} \otimes M_{c|\bar{z}^*}\otimes I_A \rho_{E_1E_2BCA}],  \quad \forall b,c,e', \\
		\sum_{l} {\rm Tr}_{E_1E_2BC}[M_{e}^1\otimes M_{c}^2\otimes M_{b|\bar{y}^*} \otimes M_{l|z^*}\otimes I_A \rho_{E_1E_2BCA}] = \sum_{l}{\rm Tr}_{E_1E_2BC}[M_{e}^1\otimes M_l^2\otimes M_{b|\bar{y}^*} \otimes M_{c|z^*}\otimes I_A \rho_{E_1E_2BCA}],  \quad \forall b,c,e.
	\end{aligned}
\end{equation}
Then, the observed assemblage can be written as
\begin{equation}
	\begin{aligned}
		\sigma_{bc|{y}^*{z}^*}^{obs} &= {\rm Tr}_{E_1E_2BC}[M_{b}^1\otimes M_{c}^2 \otimes I_B \otimes I_C \otimes I_A \rho_{E_1E_2BCA}],  \quad \forall b,c,\\
		\sigma_{bc|y^*\bar{z}^*}^{obs} &= {\rm Tr}_{E_1E_2BC}[M_{b}^1\otimes I_{E_2} \otimes I_B \otimes M_{c|\bar{z}^*}\otimes I_A \rho_{E_1E_2BCA}],  \quad \forall b,c, \\
		\sigma_{bc|\bar{y}^*{z}^*}^{obs} &= {\rm Tr}_{E_1E_2BC}[I_{E_1}\otimes M_{c}^2\otimes M_{b|\bar{y}^*} \otimes I_C\otimes I_A \rho_{E_1E_2BCA}],  \quad \forall b,c, \\
		\sigma_{bc|\bar{y}^*\bar{z}^*}^{obs} &= {\rm Tr}_{E_1E_2BC}[I_{E_1}\otimes I_{E_2} \otimes M_{b|\bar{y}^*} \otimes M_{c|\bar{z}^*} \otimes I_A \rho_{E_1E_2BCA}],  \quad \forall b,c.
	\end{aligned}
\end{equation}
Since the measurements $\{M_b^1\otimes I_B\}_b$ and $\{I_{E_1}\otimes M_{b|\bar{y}^*}\}_b$ must be compatible, and so as $\{M_c^2\otimes I_C\}_c$ and $\{I_{E_2}\otimes M_{c|\bar{z}^*}\}_c$. Hence, the observed assemblage is unsteerable in a multipartite scenario. However, in the case of $m_B\geq 3$ or $m_C\geq 3$, the additional measurements could express steerability with measurements other than $y^*$ or $z^*$ but do not be involved in generating the randomness. 

\setcounter{equation}{0}
\renewcommand\theequation{B\arabic{equation}}

\section{Appendix B: Quantification of certified Randomness in multipartite scenario}
First of all, we simplify the problem Eq. (2) in the main text to the following problem based on the classical-quantum state.
 \begin{eqnarray}
	\begin{aligned}
	\label{suppeq:18}
	&\max~~~\sum_{e, e^\prime}\operatorname{Tr}\left[  \left( I_A \otimes M_{b = e|y^*} \otimes M_{c = e^\prime|z^*} \right)  \sigma_{ABC}^{ee^\prime} \right]\\
	&{\rm w.r.t.}~~~ \{\sigma_{ ABC}^{ee^\prime} \}_{e,e^\prime},~\{M_{b|y}\}_{b,y},~\{M_{c|z}\}_{c,z}\\
	&~\operatorname{ s.t.}\quad\sum_{e,e^\prime} \operatorname{Tr}_{ BC} \left[\left(I_A \otimes M_{b |y} \otimes M_{c |z} \right) \sigma_{ABC}^{ee^\prime} \right] = \sigma_{b c|y z}^{obs},~\forall b,c,y,z, \\
	&~~~~~~~~\quad \sigma_{ABC}^{e e^\prime} \geq 0 ,\quad \forall e, e^\prime, \quad \sum_{e,e^\prime}\operatorname{Tr}\left[\sigma_{ABC}^{ee^\prime}\right] = 1,\\
	& \qquad\quad\left\{M_{b|y}\right\}_{b}, \left\{M_{c|z}\right\}_{c}\in {\rm POVM}, \quad \forall y,z.
	\end{aligned}
\end{eqnarray}
Here we don't have to fix the dimension of Eve during the calculations of the upper bound of randomness.

\subsection{1.~Randomness certified on Bob and Charlie together}\label{suppsec:BI}
\paragraph{Lower Bounds.-- } A lower bound $H_{\min}^{NS}$ can be calculated with the no-signaling constraint:
\begin{align} 
		\label{suppeq:2}
		& \max_{\{\sigma_{b c\mid y z}^{e e^\prime}\}_{e,e',b,c,y,z}} \operatorname{Tr}\left[\sum_{e, e^\prime}\sigma_{b=e,c=e^\prime|y^*z^*}^{e e^\prime}\right] \notag \\ 
		\operatorname{s.t.} \quad &\sum_{e, e^\prime}\sigma_{b c|y z}^{e e^\prime} = \sigma^{obs}_{b c|y z},    \quad \quad \quad \forall b,c,y,z, \notag \\ 
		\quad & \sigma_{b c|y z}^{e e^\prime}\geq 0, \qquad \qquad \qquad  \forall e, e^\prime,b,c,y,z, \\
		\quad &\sum_{b}\sigma_{b c|y z}^{e e^\prime} = \sum_{b} \sigma_{b c|y^\prime z}^{e e^\prime}, \quad \forall e, e^\prime,c,z ,y,y^\prime, \notag \\ 
		\quad& \sum_{c}\sigma_{b c|y z}^{e e^\prime} = \sum_{c} \sigma_{b c|y z^\prime}^{e e^\prime}, \quad \forall e, e^\prime,b,y,z ,z^\prime.\notag
\end{align}
For each $e, e^\prime$, when the optimal solution $\sigma_{bc|yz}^{e e^\prime}$ can be written as
\begin{eqnarray}
	\begin{aligned} 
		\sigma_{bc|yz}^{e e^\prime} = \operatorname{Tr}_{BCE} \left[ I_A \otimes M_{b|y}\otimes M_{c|z} \otimes M_{e e^\prime} \rho_{ABCE}\right], \quad \forall e, e^\prime,
	\end{aligned}
\end{eqnarray}
the optimal $P_g$ solved in Eq.~\eqref{suppeq:2} is equivalent to that in Eq.~(2) in the main text. Unfortunately, in the tripartite scenario, the constraint with no-signaling principle do not equivalent to the quantum realization constraint. With the NPA hierarchy~\cite{prl_2007_npa,njp_2008_npa}, we can obtain a series of tighter lower bounds $H_{\min}^{Q_k}$ by considering the constraint $\{\sigma_{bc|yz}^{ee^\prime}\}_{b,c,y,z}\in Q_k$ for each $e, e^\prime$ and a positive interger $k$. 

\paragraph{Upper Bound (see-saw Algorithm).-- } The upper bound of the min-entropy can be calculated by fixing the dimension of each subsystem. Here we use $d_A,d_B,d_C$ to denote the dimensions of Alice, Bob and Charlie, respectively. However, even by fixing the dimensions, the problem Eq.~\eqref{suppeq:18} is neither an SDP problem nor a linear problem. Here we use the see-saw algorithm to convert this problem into three SDPs. 

Firstly, we set the POVMs $\{ M_{b|y} \}_{b,y}$ and ensemble $\{ \sigma_{ABC}^{e e^\prime} \}_{e e^\prime}$ by random. In order to get matrices $ M_{b|y}^{(0)} \geq 0$, we take $n_B m_B d_B^2$ real numbers between -1 and 1 from uniform distribution to write $n_B m_B$ tridiagonal matrices $\{T_{b|y}\}_{b,y}$. Then, we have $n_B m_B$ random hermitian and positive semi-definite matrices $\{M_{b|y}^{(0)} = T_{b|y}^{\dagger} T_{b|y} / {\rm Tr} [  T_{b|y}^{\dagger} T_{b|y}]  \}_{b,y}$. We note that the set of matrices may not be satisfy with $\sum_{b} M_{b|y}^{(0)} = I$. But in the following steps, we set this condition as a constraint in SDP, then the
feasible set is still constrained by POVM condition, i.e., the final optimal solution $\{M_{b|y}\}_{b,y}$ are POVMs. For the set of matrices $\{\sigma_{ABC}^{e e^\prime(0)}\}_{e, e^\prime}$, we take $2 n_Bn_C d_A d_B d_C$ real numbers between -1 and 1 from uniform distribution to generate the amplitude terms as well as phase terms of $n_Bn_C$ pure states $\{\rho_{ABC}^{e e^\prime(0)}\}_{e,e^\prime}$. We also take $n_Bn_C$ positive numbers between 0 and 1 from uniform distribution to generate probability $p(e,e^\prime)$ with normalization $\sum_{e,e^\prime=0}^{n_Bn_C-1} p(e,e^\prime)=1$. Hence, the random state $\sigma_{ABC}^{e e^\prime(0)} = p(e,e^\prime)\rho_{ABC}^{e e^\prime(0)}$ is obtained. After giving the initial POVMs $\{ M_{b|y}^{(0)} \}_{b,y}$ and ensemble $\{ \sigma_{ABC}^{e e^\prime(0)} \}_{e, e^\prime}$, we can solve the following SDP:
	\begin{eqnarray}
		\begin{aligned}
			\label{suppeq:4}
			&\max_{\substack{\{M_{c|z}\}_{c,z}\\ \left\{\lambda_{bc|yz}\right\}_{b,c,y,z}}}\sum_{e,e^\prime}\operatorname{Tr}\left[  \left(  M_{b = e|y^*}^{(0)} \otimes M_{c = e^\prime|z^*} \right)  \sigma_{ BC}^{ee^\prime(0)}  \right]  - \mu\sum_{b,c,y,z} \lambda_{bc|yz} \\ 
			\operatorname{ s.t.} \qquad& -\lambda_{bc|yz} I\leq \sum_{ee^\prime} \operatorname{Tr}_{ BC} \left[{I}_A \otimes \left(M_{b |y}^{(0)} \otimes M_{c |z} \right) \sigma_{ABC}^{ee^\prime (0)}\right] - \sigma_{b c|y z}^{obs}  \leq \lambda_{bc|yz} I, \quad \forall b,c,y,z.\\
			&\qquad \qquad \left\{M_{c|z}\right\}_c\in {\rm POVM}, \quad \forall z, \quad  \lambda_{bc|yz}\geq 0, \quad \forall b,c,y,z,
		\end{aligned}
	\end{eqnarray}
where $\sigma_{BC}^{e e^\prime (0)} = {\rm Tr}_{A} [ \sigma_{ABC}^{e e^\prime (0)}]$. In this step, an optimal solution $\{M_{c|z}^{(1)}\}_{c,z}$ can be found easily.
Here we use a penalty to change the first equality constraint in Eq.~\eqref{suppeq:18} to an inequality constraint, where $\mu$ is about 1e2$\sim$1e3. This is because the first constraint in Eq.~\eqref{suppeq:18} could not be satisfied easily when the POVMs $\{ M_{b|y}^{(0)} \}_{b,y}$ and states $\{ \sigma_{ABC}^{e e^\prime(0)} \}_{e e^\prime}$ are set by random. Then Eq. \eqref{suppeq:4} will probably be infeasible without penalty.

In the second step, we fix the optimal solution $\{M_{c|z}^{(1)}\}_{c,z}$ and also the initial  states $\{ \sigma_{ABC}^{ee^\prime(0)} \}_{e, e^\prime}$. Then we can also find an optimal solution $\{M_{b|y}^{(1)}\}_{b,y}$ by changing the variables in Eq.~\eqref{suppeq:4}. Now the optimal solution $\{M_{b|y}^{(1)}\}_{b,y}$ is already a POVM and can be easily found in this step. 

In the third step, we fix POVMs $\left\{M_{b|y}^{(1)}\right\}_{b,y}$ and $\left\{M_{c|z}^{(1)}\right\}_{c,z}$ to find an optimal set of states $\left\{ \sigma_{ABC}^{ee^\prime(0)} \right\}_{e,e^\prime}$ that maximize Eq.~\eqref{suppeq:4}. Note that the third step will take much time than the first two steps, so we iterate the first two steps until the guessing probability reaches its convergency. 

We iterate these steps until $\max\{\lambda_{bc|yz}\}$ decreases to about 1e-9 and the optimal guessing probability reaches its convergency, hence an upper bound $H_{\min}^{Dim}$ can be calculated. Furthermore, when we find that the guessing probability changes with a very slow rate, adding some random POVMs and states (with tiny weight) is helpful for finding a larger $P_g$. We note that the solution we found by see-saw algorithm may not be a global optimal solution, which means the optimal $P_g$ we got is less than the global solution. So the solution in this part is still an upper bound of of actual randomness. 

\subsection{2.~Randomness certified on Bob or Charlie solely} 
In a multipartite scenario, we can also consider that Eve only guess the measurement results of Bob or Charlie solely. Here we give a definition of the randomness gernerated from only Bob, and it is the same with Charlie. In this case, the guessing probability $P_g^B$ is given by 

\begin{eqnarray}
	\begin{aligned}
		\label{suppeq:5}
		&\max_{\substack{\{\sigma_{\rm ABC}^{e} \}_{e}\\\{M_{b|y}\}_{b,y}\\\{M_{c|z}\}_{c,z}\\}}\sum_{e}\operatorname{Tr}\left[  M_{b = e|y^*}   \sigma_{ B}^{e} \right]\\ 
		\operatorname{ s.t.} &\sum_{e} \operatorname{Tr}_{BC} \left[\left( {I}_A \otimes M_{b |y} \otimes M_{c |z} \right) \sigma_{ ABC}^{e} \right] = \sigma_{b c|y z}^{obs}, \quad \forall b,c,y,z, \\
		&\quad \sigma_{ABC}^{e} \geq 0 , \quad \forall e, \quad \quad \sum_{e}\operatorname{Tr}\left[\sigma_{ABC}^{e}\right] = 1,\\
		&\quad \left\{M_{b|y}\right\}_{b,y}, \left\{M_{c|z}\right\}_{c,z}\in {\rm POVM},
	\end{aligned}
\end{eqnarray}
where $\sigma_{B}^e = {\rm Tr}_{AC} [\sigma_{ABC}^e]$. The main difference between Eq.~\eqref{suppeq:5} and Eq.~\ref{suppeq:18} is just the objective function, which means we certify the randomness on Bob still in a real tripartite scenario, which is different from the previous bipartite scenario. So we still can not solve the above optimization problem directly.

For the lower bounds of Eq.~\eqref{suppeq:5} , it can also be calculated by utilizing the no-signaling principle:
\begin{eqnarray}
	\begin{aligned} 
		\label{eq:app1}
		& \max_{\{\sigma_{b c\mid y z}^{e}\}_{e,b,c,y,z}} \operatorname{Tr}\left[\sum_{e,c }\sigma_{b=e,c|y^*0}^{e}\right]\\ 
		\operatorname{s.t.} \quad &\sum_{e}\sigma_{b c|y z}^{e} = \sigma^{obs}_{b c|y z},    \quad \quad \quad \forall b,c,y,z,  \\ 
		\quad & \sigma_{b c|y z}^{e}\geq 0, \qquad \qquad \qquad  \forall e,b,c,y,z, \\
		\quad &\sum_{b}\sigma_{b c|y z}^{e} = \sum_{b} \sigma_{b c|y^\prime z}^{e }, \quad \forall e,c,z ,y,y^\prime,\\ 
		\quad & \sum_{c}\sigma_{b c|y z}^{e } = \sum_{c} \sigma_{b c|y z^\prime}^{e }, \quad \forall e,b,y,z ,z^\prime.
	\end{aligned}
\end{eqnarray}
Also some tighter lower bounds can be found by adding NPA hierarchy constraints $\{\sigma_{bc|yz}^e\}_{b,c,y,z}\in Q_k$, $\forall e$.

For the upper bound, we can also use see-saw algorithm to calculate the upper bound $H_{\min}^{Dim}$ in this scenario when we fix the dimension of each subsystems. The steps are same with the method presented in section \ref{suppsec:BI}. 

\setcounter{equation}{0}
\renewcommand\theequation{C\arabic{equation}}

\section{Appendix C:~Example of continuous-variable case}

Here we give an example to certify randomness in the continuous-variable system. The pure state $|\Psi\rangle_{CV}$ generated from the setup shown in Fig.~4(a) in the main text is
\begin{align}
    &\hat{U}_{T_2}^{B C}\hat{U}_{\eta_1}^{A B}\hat{S}(\xi_A)\hat{S}(\xi_B)|0_A\rangle|0_B\rangle|0_C\rangle
    = \sum_{p=0}^{\infty}\sum_{k_1= 0}^p \sum_{k_2 = 0} ^{k_1} \sum_{k_3 = 0} ^{p-(2k_2-k_1)}\frac{(\tanh r)^{p}}{2^{p} \cosh r} \left(4\sqrt{\eta_1\left(1-\eta_1\right)} \right)^{p-k_1}  \left(2\eta_1 - 1 \right)^{k_1}(-1)^{p-(3k_2+k_3)}    \\
    & \frac{\sqrt{\left(p+\left(2k_2-k_1\right)\right)!\left(p-\left(2k_2-k_1\right)\right)!}}{\left(p-k_1\right)!\left(k_1-k_2\right)! k_2!}\sqrt{C_{p-(2k_2-k_1)}^{k_3}\frac{\eta_2^{k_3}}{(1-\eta_2)^{k_3-({p-(2k_2-k_1)})}}}|p + \left(2k_2-k_1\right)\rangle_A|k_3\rangle_B|{p-(2k_2-k_1+k_3)}\rangle_C \notag
\end{align}
where $\hat{S}(\xi)=e^{\frac{1}{2}\left(\xi^{*} \hat{a}^{2}-\xi\left(\hat{a}^{\dagger}\right)^{2}\right)}$ is the squeezed operator and $\xi = re^{i\theta}$.

In order to generate the assemblage measured by Alice, we bin the homodyne measurement results on Bob and Charlie by the coarse-graining scheme~\cite{pra_2022_the,prl_2018_mubcv}. For example, the measurements on Bob can be written as:
\begin{eqnarray}
	\begin{aligned}
		{M}_{b \mid y}=\int_{\mathbb{R}} f_b\left(z, T_{y }\right)|z\rangle_{y}\langle z| d z,
	\end{aligned}
\end{eqnarray}
where $y = \hat x, \hat p$ is the input, $T_y$ is the period, and $f_b(z,T_{y})$ is a function:
\begin{eqnarray}
	f_{b}\left(z, T_{y}\right)= \begin{cases}
		1, & b s_{y} \leq z \bmod T_{y}<(b+1) s_{y} \\ 0, & \text { otherwise }
	\end{cases}
\end{eqnarray}
where $s_{y} = T_{y}/n_B$ is the width of the bins and $n_B$ is the number of outcomes, i.e. $b\in\{0,1,\cdots,n_B-1\}$. Here we set $T_{\hat p} = 2\pi n_B/T_{\hat x}$ to ensure mutual unbiasedness. So based on such POVMs, the assemblage measured by Alice is $\sigma_{bc|yz}^{obs} = {\rm Tr}_{BC}[I_A\otimes M_{b|y}\otimes M_{c|z} |\Psi\rangle_{CV}\langle \Psi|]$.

In Fig.~4 in the main text, we set $r = 3$dB and fix the transmissivity of the first beam splitter,  to $\eta_1 = 1/2$. The state $|\Psi\rangle_{CV}$ changes with the second beam splitter transmissivity $\eta_2$. Then by analyzing the assemblage on Alice, we can calculate an upper bound and some lower bounds of $H_{\min}$ on Bob's and Charlie's measurement results ($y^*=z^*=\hat x$). Here we cut off the Fock basis to one photon which is enough in our case, although higher ``cut off basis" would give higher $H_{\min}^{NS}$. Note that $H_{\min}^{NS}$ is maximised over binning periods in the range $T_{\hat x} (Bob)\in [2,10]$ for Bob and $T_{\hat x} (Charlie) \in [2,10]$ for Charlie independently. Then $H_{\min}^{Q_1}$ is calculated in these optimized periods $T_{\hat x}(Bob)$ and $T_{\hat x}(Charlie)$. Moreover, we also give the randomness on only Bob or only Charlie based on the same steps.

\setcounter{equation}{0}
\renewcommand\theequation{D\arabic{equation}}
\renewcommand\thefigure{D\arabic{figure}}
\section{Appendix D:~Randomness Certified by A Steering Inequality Violation}
%We have presented a method to certify the randomness by analyzing the assemblage $\left\{\sigma_{bc|yz}^{obs}\right\}_{b,c,y,z}$ observed by Alice. However, reconstructing conditional states by Quantum State Tomography (QST) requires a large amount of measurements. To avoid doing this, we extend the analyse to the case that Alice also do some known POVMs $\left\{M_{a|x}\right\}_{a,x}$ ~\cite{prl_2018_paul,pra_2022_the,prl_2019_shumingcheng}, then we can certify randomness from a joint probability distribution $p^{obs}(abc|xyz)$. Thus, the first constraint in Eq. 2 in the main text can be converted to
%\begin{eqnarray}
	%\begin{aligned}
		%{\rm Tr}  \left[M_{a|x}\otimes M_{b|y}\otimes M_{c|z} \otimes I_E \rho_{\rm ABCE}\right] = P^{obs}(abc|xyz),  \qquad \forall a,b,c,x,y,z,
	%\end{aligned}
%\end{eqnarray}
%where the POVMs done by Alice (trusted side) are known. 

The randomness on Bob's and Charlie's measurement results can also be certified by observing a violation of steering inequality $V$, which can be solved by the following optimisation problem:
\begin{eqnarray}
	\begin{aligned}
		\label{suppeq:8}
		&\max_{\substack{\{\sigma_{ABC}^{ee^\prime} \}_{e,e^\prime}\\\{M_{b|y}\}_{b,y}\\\{M_{c|z}\}_{c,z}\\}}~~~\sum_{e,e^\prime}\operatorname{Tr}\left[  \left(  M_{b = e|y^*} \otimes M_{c = e^\prime|z^*} \right)  \sigma_{BC}^{ee^\prime} \right]\\
		&\quad \operatorname{ s.t.}\quad \sum_{\substack{a,b,c,\\ x,y,z\\ee'}}F_{abc|xyz} {\rm Tr}  \left[M_{a|x}\otimes M_{b|y}\otimes M_{c|z}\sigma_{ABC}^{ee'}\right] = V, \\
		&\qquad  ~~~\quad \sigma_{ABC}^{ee^\prime} \geq 0, \quad \forall e,e',\quad \sum_{ee^\prime}\operatorname{Tr}\left[\sigma_{ABC}^{ee^\prime}\right] = 1,\\
		& \qquad\quad ~~\left\{M_{b|y}\right\}_{b,y}, \left\{M_{c|z}\right\}_{c,z}\in {\rm POVM},
	\end{aligned}
\end{eqnarray}
where $\sigma_{BC}^{ee^\prime} = {\rm Tr}_{A} [\sigma_{ABC}^{ee^\prime}]$, $F_{abc|xyz}$ are the coefficients of the given inequality. Note that it is still considered in a real tripartite scenario, so the actual min-entropy need to be described by its lower and upper bounds. Also it can be easily extended to the case that randomness only on Bob or Charlie solely, and the method is similar with Appendix~B.%section~\ref{suppsec:B}.

\setcounter{figure}{0}
\begin{figure}[h]
	\includegraphics[width=0.5\textwidth]{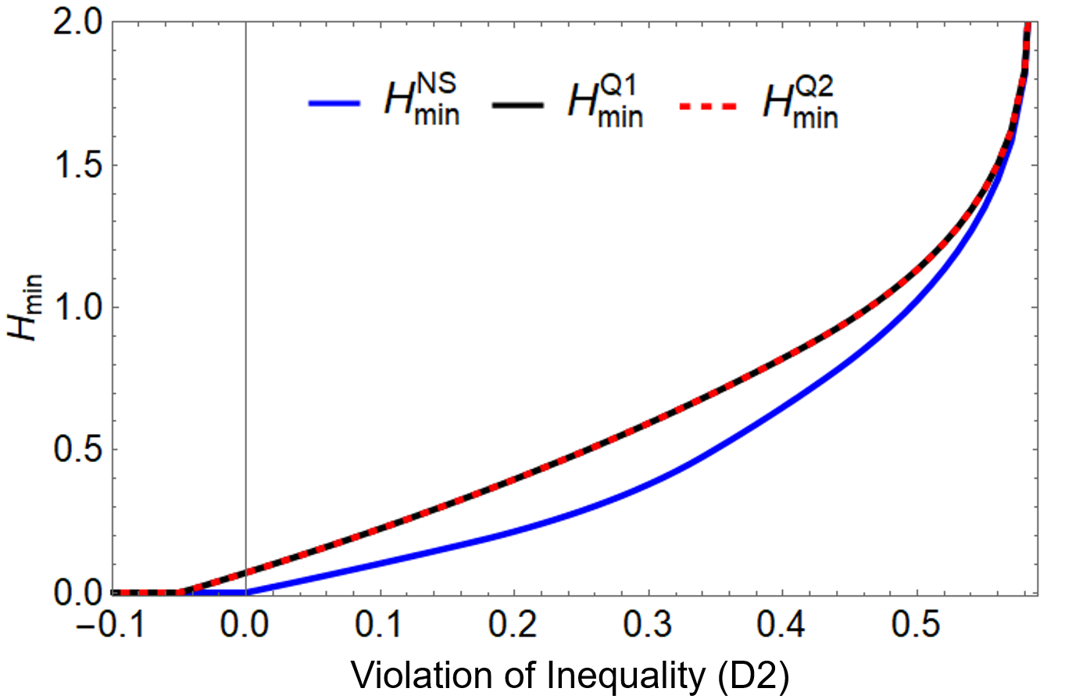}
	\caption{\label{suppfig:vio} \textbf{Randomness certified by observing a violation of genuine tripartite steering ineqaulity.}}
\end{figure}

For example, when Bob and Charlie are untrusted, a genuine tripartite steering can be observed if \cite{nc_2015_multipartite}
\begin{eqnarray}
	\begin{aligned}
		\label{suppeq:9}
		\alpha\left\langle A_{3} B_{3}\right\rangle+\alpha\left\langle A_{3} Z\right\rangle+\alpha\left\langle B_{3} Z\right\rangle+\beta\left\langle A_{1} B_{1} X\right\rangle -\beta\left\langle A_{1} B_{2} Y\right\rangle-\beta\left\langle A_{2} B_{1} Y\right\rangle-\beta\left\langle A_{2} B_{2} X\right\rangle-1 \leq 0
	\end{aligned}
\end{eqnarray}
is violated, in which $\alpha = 0.1831$ and $\beta = 0.2582$. A violation of Eq.~\eqref{suppeq:9} implies the assemblage could not arise from measurements on a bi-separable state $\rho^{\mathrm{bisep}} =\sum_\lambda p_\lambda^{\mathrm{A}: \mathrm{BC}} \rho_\lambda^{\mathrm{A}} \otimes \rho_\lambda^{\mathrm{BC}}+\sum_\lambda p_\lambda^{\mathrm{B}: \mathrm{AC}} \rho_\lambda^{\mathrm{B}} \otimes \rho_\lambda^{\mathrm{AC}} +\sum_\lambda p_\lambda^{\mathrm{AB}: \mathrm{C}} \rho_\lambda^{\mathrm{AB}} \otimes \rho_\lambda^C$~\cite{nc_2015_multipartite}. 
Taking it into Eq.~\eqref{suppeq:8}, the randomness bounded by a genuine tripartite steering inequality violation is shown in Fig.~\ref{suppfig:vio}.

We can find an obvious distinction between the curves of $H_{\min}^{NS}$ and $H_{\min}^{Q_k}$. This is reasonable because there are many different assemblages corresponds with a same violation. Thus, the difference between the no-signaling set and the quantum set widens, resulting in a significant difference in randomness. Furthermore, because genuine multipartite steering is not necessary for randomness, there is extra randomness can be certified without violating Eq.~\eqref{suppeq:9}.

	\bibliography{multipartiterandomness_Ref}

\begin{thebibliography}{70}
\bibitem [{foo()}]{footnote}%
  \BibitemOpen
  \href@noop {} {}\bibinfo {note} {One may think the distinction between
  $H_{\min}^{NS}$ and $H_{\min}^{Q_k}$ is tiny. This is because we certify
  randomness by an assemblage for some given quantum states. We also analyze
  the certified multipartite randomness when only a violation of steering
  inequality is observed via a joint probability distribution
  $p^{obs}(abc|xyz)$ (See Appendix~D), which shows a non-ignorable distinction
  between different lower bounds.}\BibitemShut {Stop}%
\end{thebibliography}

\end{document}